\documentclass[prl,twocolumn,showpacs]{revtex4}

\usepackage{amsmath}
\usepackage{amsfonts}
\usepackage{amsthm}
\usepackage{amssymb}
\usepackage{amscd}
\usepackage[british]{babel}
\usepackage{graphicx}
\usepackage{psfrag}
\usepackage{epsfig}
\usepackage{rotating}
\usepackage{times}

\theoremstyle{plain}

\newtheorem{remark}{Remark}[section]

\newcommand{\boxend}{\flushright{$\Box$}}

\begin{document}

\title{Non-singular models of  universes in teleparallel theories}

\author{Jaume de Haro$^{}$\footnote{E-mail: jaime.haro@upc.edu}  and Jaume Amoros$^{}$\footnote{E-mail: jaume.amoros@upc.edu}}

\affiliation{$^{}$Departament de Matem\`atica Aplicada I, Universitat
Polit\`ecnica de Catalunya, Diagonal 647, 08028 Barcelona, Spain}

%
%


\thispagestyle{empty}

\begin{abstract}
Different models of universes are considered in the context of teleparallel theories. Assuming that the universe is filled by a fluid with
equation of state (EoS) $P=-\rho-f(\rho)$, for different teleparallel theories and different EoS
we study its dynamics.
Two particular cases are studied in detail: in the first one we consider a function
 $f$ with two zeros (two de Sitter solutions) that mimics a huge cosmological constant at early times and a
pressureless fluid at late times. In the second one, in the context of loop quantum cosmology
with a small cosmological constant, we consider a pressureless fluid ($P=0\Leftrightarrow
f(\rho)=-\rho$) which means that there are a de Sitter and an anti de Sitter solution.
  In both cases  one
obtains  a non-singular universe that at early times  is in an inflationary phase, after
leaving this phase it passes trough a matter dominated phase and finally at late times it expands in an accelerated way.
\end{abstract}

\pacs{04.50.Kd, 98.80.-k, 98.80. Jk}

\maketitle

{\it 1. Introduction.}--- Teleparallel theories ($F(T)$ theories) \cite{Hehl:1976kj,Hayashi:1979qx,Flanagan:2007dc,Garecki:2010jj}
are built from the scalar torsion $T=-6H^2$, where $H$ is the Hubble parameter. 
The field equations are second-order, which  is a great advantage to $F(R)$ theories, whose fourth-order equations lead to pathologies like instabilities or
large corrections to Newton's law \cite{no07a,l10}.
This entails that
the modified Friedmann equation  depicts a curve in the plane $(H,\rho)$, that is, the universe moves along this curve an
its dynamics is given by the modified Raychaudhuri equation and the conservation equation.

This opens the possibility to built non-singular models of universes filled by a fluid with equation of state (EoS) $P=-\rho-f(\rho)$, being $P$ the
pressure and $\rho$ the energy
density. For this EoS, the zeros of the function $f$ give de Sitter and anti de Sitter solutions. Then, choosing a function $f$ with two zeros one obtains a non-singular
model of universe.

As examples of teleparallel theories we study Einstein Cosmology (EC) with and without a small cosmological constant, loop quantum cosmology (LQC)
with and without a small cosmological constant,
 and the teleparallel version of the model $F(R)=\frac{R}{2}-\frac{\mu^4}{2R}$,
$R$ being the scalar curvature.
For these teleparallel examples we consider  models of EoS, which mimic a huge cosmological constant at early times and a
pressureless fluid at late times, like $P=-\frac{\rho^2}{\rho_i}$,
$\rho_i$ being the initial energy density of the universe.

 We will see how these simple models could solve the so-called coincidence problem due to periods of acceleration of our universe, that is, they
  could mimic the evolution of a universe that begins in an inflationary phase, after leaving it passes trough a matter dominated one, and
at late times enters in an accelerated phase.

We also study  the specific case of a universe  filled by
a pressureless fluid ($P=0$) in the context of LQC with a small cosmological constant. In that case, at early times the universe is in an anti de Sitter phase, after leaving this phase it starts to
accelerate (inflationary phase) leaving the   contracting phase to enter in the expanding one (it bounces),
then it starts to decelerate and enters in a matter dominated phase. Finally, at late times it enters in a
 de Sitter phase
(late time cosmic acceleration).

In some sense our work seems a teleparallel version of the early idea that $F(R)$ cosmology
could unify inflation with late cosmic acceleration \cite{no03} if one uses a function $F$ of the form
$F(R)=\frac{1}{2}\left(R-\frac{a}{(R-\Lambda_1)^n}+b(R-\Lambda_2)^m \right)$, with $n,m,a,b,\Lambda_1,\Lambda_2\geq 0$ parameters.
However,
our point of view is different from the one adopted
in current dark energy cosmology (see \cite{bcno12} for a recent review):  Where,
in models of oscillating dark energy  one takes an EoS of the form
$P=\omega(t)\rho$ (see for example \cite{no06,flpz06}), or a scalar field under the action of an  oscillating potential \cite{dks00}.
It is also different from the current $F(T)$ or $F(R)$ theories  where, in order to reconstruct models realizing cosmologies,
the authors have to use very complicated functions $F$ (see for example \cite{bmno12, no07,cenosz08,ccdds11}).

The units  used in the paper are $ c = \hbar = 8\pi G=1$.
\medskip

{\it 2. Teleparallel cosmological theories.}---Teleparallel theories are based in the Weitzenb\"ock space-time. To built this space-time one  chooses  a global system of four orthonormal
vector fields $\{{\bf e}_i\}$ which are related with the vectors
$\{\partial_{\mu}\}$ via the relation ${\bf e}_i=e_i^{\mu}\partial_{\mu}$. Then one introduces a covariant derivative $\nabla$ that defines
absolute parallelism with respect the global basis $\{{\bf e}_i\}$, that is $\nabla_{\nu}e_i^{\mu}=0$. From this, one obtains the metric
Weitzenb\"ock connection $\Gamma^{\gamma}_{\quad\mu \nu}=e_i^{\gamma}\partial_{\nu}e^i_{\mu}$. (Note that this connection is metric, it
satisfies $\nabla_{\gamma}g_{\mu\nu}=0$).

The Weitzenb\"ock space-time has identically vanishing  curvature (the Riemann tensor vanishes), but it is not torsion free. One effectively has
$T^{\gamma}_{\quad\mu \nu}=\Gamma^{\gamma}_{\quad \nu\mu}-\Gamma^{\gamma}_{\quad\mu \nu}\not=0$.
Then in order to built the Lagrangian in the  Weitzenb\"ock space-time, one has to introduce the contorsion tensor:
\begin{eqnarray}\label{1}
K^{\mu\nu}_{\quad \gamma}=-\frac{1}{2}\left(T^{\mu\nu}_{\quad \gamma}- T^{\nu\mu}_{\quad \gamma}-
T^{\quad\mu\nu}_{\gamma}
 \right),\end{eqnarray}
and the tensor
\begin{eqnarray}\label{2} S^{\quad\mu\nu}_{\gamma}=\frac{1}{2}\left(K^{\mu\nu}_{\quad \gamma}+\delta^{\mu}_{\gamma}T^{\theta\nu}_{\quad \theta}-
\delta^{\nu}_{\gamma}T^{\theta\mu}_{\quad \theta}
   \right),\end{eqnarray}
to construct the so-called {\it scalar torsion}
\begin{eqnarray}\label{3}
T= S^{\quad\mu\nu}_{\gamma} T^{\gamma}_{\quad\mu \nu},\end{eqnarray} and one can define modified teleparallel theories via a Lagrangian of the
form  ${\mathcal L}_{T}=VF(T)-V\rho$, where $V$ is the volume of the spatial part.

What is interesting here is that if one chooses the basis $\{{\bf e}_0=\partial_{0}, {\bf e}_1=a\partial_{1}, {\bf e}_2=a\partial_{2},
 {\bf e}_3=a\partial_{3}\}$, then for the Friedmann-Lema{\^\i}tre-Robertson-Walker (FLRW) metric one obtains  \cite{ Bengochea:2008gz}  $T=-6H^2$.

The problematic point in teleparallelism is that
it depends on the choice of the global basis.
In the sense that, if one uses a local
Lorentz transformation to transform the original  global basis in another one,
in general, one obtains another Weitzenb\"ock connection and thus $T$ could
change~\cite{lsb11, Li:2011rn}, which does not happen in
modified gravity ($F(R)$ theories) where the invariants do not depend of a global basis.
However, for a FLRW metric, if we use
local Lorentz transformations that only depend on the time, that is,
of the form ${\bf \bar{e}}_i=\Lambda^k_i(t){\bf e}_k$,
then, even though the torsion tensor changes, the torsion scalar $T$ remains
constant with a value of $T=-6H^2$.
{}From our point of view, this gives a consistency
with teleparallel theories in
cosmology, because
$T$ is invariant from ``isotropic and homogeneous''
local Lorentz transformations.


{\it 3. General features in teleparallel theories.}---
For the flat FLRW space-time  filled by a perfect fluid with  equation of state (EoS) $P(\rho)=-\rho-f(\rho)$, teleparallel theories,
are obtained from
the Lagrangian
\begin{equation}\label{4}{\mathcal L}=VF(T)-V\rho(V),
\end{equation}
 where $\rho(V)$ is obtained from the conservation equation (first principle of thermodynamics)
 \begin{equation}\label{5}
 d(\rho V)=-PdV\Longleftrightarrow \frac{d\rho}{dV}=\frac{f(\rho)}{V}\Longleftrightarrow \dot{\rho} {=} 3Hf(\rho).
 \end{equation}

The conjugate momentum is  then given by
$p_V=\frac{\partial {\mathcal L}}{\partial\dot{V}}= -4HF'(T)$, and thus the Hamiltonian is
\begin{eqnarray}\label{6}
{\mathcal H}=
\dot{V}p_V- {\mathcal L}= (2TF'(T)-F(T)  +\rho )V.\end{eqnarray}

In general relativity  the Hamiltonian is constrained to be zero. (In $f(T)$ gravity the Hamiltonian is also constrained to be zero, this comes
from the equation given by the variation of the action with respect the vierbien $e^{\mu}_i$. Actually, the constrain comes from the $(0,0)$ component
of this equation (see eq. $(9)$ of \cite{ccdds11})).
This constrain leads to the modified
Friedmann equation
\begin{eqnarray}\label{7}
\rho=-2 F'(T)T+F(T)\equiv G(T),
\end{eqnarray}
which is a curve in the plane $(H,\rho)$.

Conversely,
given a curve of the form $\rho=G(T)$
for some function $G$, it could be obtained from the modified Friedmann
equation  by choosing
\begin{eqnarray}\label{8}
F(T)=-\frac{\sqrt{-T}}{2}\int \frac{G(T)}{T\sqrt{-T}}dT.
\end{eqnarray}

The modified Raychaudhuri equation is obtained from the modified Friedmann equation
taking its derivative with respect  to the time, giving rise to
the equation $\dot{H}=-\frac{f(\rho)}{4G'(T)}$.
Then, the dynamics of the universe is given by the modified Raychaudhuri equation and the conservation equation, i.e. by the system
\begin{eqnarray}\label{9}
\left\{\begin{array}{ccc}
\dot{H} &{=}& -\frac{f(\rho)}{4G'(T)}\\
\dot{\rho} &{=}& 3Hf(\rho),
\end{array}
\right. \end{eqnarray}
provided the universe moves along the curve $\rho=G(T)$.

What is important  to stress here is that the critical points of the system are given by the solutions of the equation $f(\rho)=0$,
that is, the de Sitter and anti de Sitter solutions are the solutions of the equation $f(\rho)=0$.
This gives us a simple way to built non-singular models of universes: One has to choose a function $f$ with two zeros (two de Sitter solutions, or
a de Sitter and an anti de Sitter solutions),
then the universe will move from one to the other.

To finish this section, we will introduce three parameters which are important in cosmology:
\begin{enumerate}
\item $\omega\equiv \frac{P}{\rho}$  is useful to classify the fluid that fills the universe (when $\omega<-1$  the fluid is called phantom, when $\omega=0$  dust,
when $\omega=1/3$ radiation, ...).
\item $\omega_{eff}\equiv \frac{1}{3}\left(1-\frac{R}{3H^2} \right)=-1-\frac{2\dot{H}}{3H^2}$ is related to the expansion of the universe. Actually, when $\omega_{eff}<-1/3$
(respectively $\omega_{eff}>-1/3$)
the universe accelerates (respectively decelerates).
\item
 $\Omega\equiv \frac{\rho}{3H^2}$ gives the amount of matter in the universe.
\end{enumerate}

For a general teleparallel theory $\rho=G(T)$ one has $\omega= -1-\frac{f(\rho)}{\rho}$ and
\begin{eqnarray}\label{10}&&
\omega_{eff}=-1-\frac{f(\rho)}{TG'(T)},
\quad
\Omega=-\frac{2\rho}{T}.
\end{eqnarray}

{\it 4. Examples of teleparallel theories.}---
In this Section we will study three important examples of teleparallel theories: Einstein cosmology, loop quantum cosmology
\cite{bho12}
and the teleparallel
version of the   $F(R)=\frac{R}{2}-\frac{\mu^4}{2R}$ model \cite{cdtt04, cfdett05}.

\subsection{Einstein cosmology}
The first example of teleparallel theory is Einstein Cosmology (EC),
which is obtained from
the Lagrangian
\begin{eqnarray}\label{11}
{\mathcal L}_E=\frac{1}{2}R V-\rho(V) V
\end{eqnarray}
 where $R=6(\dot{H}+2H^2)$ is the scalar curvature.

Note that, the Lagrangian can be written as follows ${\mathcal L}_E= 3\frac{d\dot{V}}{dt}+ \bar{\mathcal L}_E$, where
\begin{eqnarray}\label{12}
\bar{\mathcal L}_E=\frac{1}{2}T V-\rho V.
\end{eqnarray}

Then since ${\mathcal L}_E$ and $\bar{\mathcal L}_E$ differ by a total derivative, one can conclude  that they are equivalent.

In EC equation (\ref{7}) becomes the
Friedmann equation
\begin{eqnarray}\label{13}
\frac{1}{2}T+\rho=0\Longleftrightarrow
H^2=\rho/3,
\end{eqnarray}
which depicts a parabola in the plane $(H,\rho)$, that is, the evolution of the universe
follows this parabola, and its  dynamics is given by  the system
\begin{eqnarray}\label{14}\left\{\begin{array}{ccc}
 \dot{H}&=&\frac{f(\rho)}{2}\\
\dot{\rho}&=&3Hf(\rho).\end{array}\right.
\end{eqnarray}

Note
 that in EC one always has $\omega=\omega_{eff}(\rho)=-1-\frac{f(\rho)}{\rho}$ and $\Omega\equiv 1$.
Once we have obtained the dynamics in EC we can study some examples of EoS.
The simplest one is given by a fluid with $\omega=\omega_0$ constant, which means
that $f(\rho)=-(1+\omega_0)\rho$, and thus, the only critical point of the system is $(0,0)$. In that case, it is
trivial to show that for $\omega_0>-1$  the universe moves from $\rho=\infty$ to $\rho=0$ (Big Bang singularity).
And for  a phantom field
($\omega_0<-1$)  the universe moves from $\rho=0$ to $\rho=\infty$ (Big Rip singularity).

To finish EC  we will built a cosmological model without singularities. One could do this
introducing a small cosmological constant, namely $\mu^4$, which could be done changing in the Lagrangian
(\ref{12})
$\rho$ by $\rho+\mu^4$. Once we have introduced this small cosmological constant,
we will choose a fluid that for large values of $\rho$ mimics a huge cosmological constant ($\Lambda=\rho_i)$, and for small values
of $\rho$ becomes pressureless. An example of this kind of fluids is given by  an EoS
\begin{equation}\label{15}
P=-\frac{\rho^2}{\rho_i}  \Leftrightarrow  \omega=-\frac{\rho}{\rho_i} \Leftrightarrow f(\rho)=-\rho\left(1-\frac{\rho}{\rho_i}\right).
\end{equation}

Then, the system has two critical points $a_f=(\mu^2/\sqrt{3},0)$ and $a_i=(\sqrt{\mu^4+\rho_i}/\sqrt{3},\rho_i)$.
At the point $a_i$ one has $\omega=\omega_{eff}=-1$ and $\Omega\cong 1$ (de Sitter phase),  at $a_f$ one has $\omega=\omega_{eff}=-1$ and $\Omega=0$ (de Sitter phase), and when
$\mu^4\ll \rho\ll\rho_i$ one has $\omega=\omega_{eff}\cong 0$ and $\Omega\cong 1$ (matter dominated  phase).

This model shows a universe evolving from an early  inflationary phase (point $a_i$) to a late time accelerated expansion (point $a_f$)
passing trough a matter dominated phase.

\begin{remark}
Note that asymptotically, this model  converge to the LCDM model, because
at late times $\rho\rightarrow 0$ and consequently the cosmological constant $\mu^4$ dominates. Then the deceleration, the
jerk and snark parameters (see \cite{bmno12} for the definition of these parameters) converge respectively to $-1$, $1$ and $s=0$. Note also that the model
may be solved analytically and then one could evaluate these parameters in the different phases in which our universe evolves. 
\end{remark}

\subsection{Loop quantum cosmology}
The main idea of LQC is that it assumes a discrete nature of space which leads, at quantum level, to consider a Hilbert
space  where
quantum states are represented by
 almost periodic functions of the dynamical part of the connection \cite{abl03,as11,a07}.
Unfortunately, the connection variable doesn't correspond to a well defined quantum operator in this Hilbert space, what leads to
 re-express the gravitational part of the
Hamiltonian in terms
of almost periodic functions,  which could be done from a  process of regularization.
This new regularized (effective) Hamiltonian
 introduces a quadratic modification ($\rho^2$)
in the Friedmann equation
at high energies \cite{svv06,s09a}, which gives rise to a bounce when the energy density becomes equal to a critical value of the order of
the Planck energy density.
This modified Friedmann equation depicts  the following  ellipse in the plane $(H,\rho)$
\begin{eqnarray}\label{16}
H^2=\frac{\rho}{3}\left(1-\frac{\rho}{\rho_c}\right)
\Longleftrightarrow \frac{H^2}{\rho_c/12}+\frac{(\rho-\frac{\rho_c}{2})^2}{\rho_c^2/4}=1,
\end{eqnarray}
where $\rho_c\equiv \frac{3}{\gamma^2\lambda^2}$
is the so-called critical density, with
$\gamma\cong 0.2375$  being the Barbero-Immirzi parameter and $\lambda$  a
parameter with dimensions of length, which is determined invoking
the quantum nature of the geometry, that is, identifying its square with the
minimum eigenvalue of the area operator in LQG, which gives as a result $\lambda\equiv
\sqrt{\frac{\sqrt{3}}{4}\gamma}$ (see \cite{s09a}).

The dynamics is now given in LQC by
the system
\begin{eqnarray}\label{17}\left\{\begin{array}{ccc}
 \dot{H}&=&\frac{f(\rho)}{2}\left(1-\frac{2\rho}{\rho_c} \right)\\
\dot{\rho}&=&3Hf(\rho),\end{array}\right.
\end{eqnarray}
where the first equation is the modified Raychaudhuri equation in LQC.
The parameters $\omega_{eff}$ and $\Omega$  become
\begin{equation}\label{18}
  \omega_{eff}=-1-\frac{f(\rho)}{\rho}\frac{\rho_c-2\rho}{\rho_c-\rho},\quad
\Omega=\frac{\rho_c}{\rho_c-\rho}.
\end{equation}

As a first example, we can study an EoS with $f(\rho)<0$ (non-phantom fluid because $\omega>-1$).
Then one obtains a cyclic universe moving in an clockwise sense along the ellipse. This universe bounces at
 points $a_1=(0,0)$  and $a_2=(0,\rho_c)$. In $a_1$ the universe enters in the contracting phase and
in $a_2$ the
universe enters in the expanding one.
On the other hand, if one
considers a phantom fluid $f(\rho)>0$, one obtains a cyclic universe moving in an anti-clockwise sense along the ellipse.

As a second example, we consider a fluid with $\omega=\omega_0$ constant, which means that $f(\rho)=-(1+\omega_0)\rho$, and thus, the only
 critical point of the system is $(0,0)$. In this situation, it is
easy to show that for $\omega_0>-1$  the universe moves from $(0,0)$ to itself in a clockwise sense, and for a phantom fluid
($\omega_0<-1$)  the universe moves from $(0,0)$ to itself in an anti-clockwise sense.

Third example is more interesting because we will build  a model  without singularities.
To do this
we perform a small modification in the modified Friedmann equation, introducing a small
cosmological constant $\mu^4$ (changing $\rho$ by $\rho+\mu^4$ in equation (\ref{16})) satisfying $\mu^4\ll \rho_c$. Then, equation (\ref{16}) becomes the ellipse
\begin{eqnarray}\label{19}
H^2=\frac{\rho+\mu^4}{3}\left(1-\frac{\rho+\mu^4}{\rho_c}\right).
\end{eqnarray}

For a pressureless fluid ($P=0\Leftrightarrow f(\rho)=-\rho$), the system has two critical points
$a_f\equiv (\frac{\mu^2}{\sqrt{3}}\sqrt{\left(1-\frac{\mu^4}{\rho_c}\right)},0)$ and
$a_i\equiv(-\frac{\mu^2}{\sqrt{3}}\sqrt{\left(1-\frac{\mu^4}{\rho_c}\right)},0)$. The first one is a de Sitter solution and the second one is
 an anti de Sitter solution.
In that case the universe moves along the ellipse from $a_i$ to $a_f$ in a clockwise sense. When it arrives at the point
$a_1\cong (-\rho_c/4,\rho_c/4)$ the universe starts to accelerate (inflationary phase) because $\frac{\ddot{a}}{a}=\dot{H}+H^2\geq 0$
when $\rho\gtrsim \rho_c/4$,
then it bounces at the top of the ellipse, i.e., at the point $a_2=(\rho_c-\mu^4,0)$ the universe leaves the contracting phase and enters the
expanding one where the energy density starts to decrease. When the universe  arrives at $a_3\cong (\rho_c/4,\rho_c/4)$ it starts to
decelerate and when the density satisfies $\rho_c\ll \rho\ll \mu^4$ the universe enters in a matter dominated phase ($\omega_{eff}\cong 0$ and
 $\Omega\cong 1$). Finally, after leaving this phase it goes asymptotically, at late times, to the point $a_f$  (de Sitter phase that mimics the late time
accelerated cosmic expansion).

Note that,  a model of universe moving from a inflationary phase towards a late time accelerating phase passing trough a matter
dominated
phase, could also obtained in LQC with a small cosmological constant by choosing as EoS of the model given in (\ref{15}), with $\rho\approx\rho_c/2$.
For this EoS, the universe moves from the de Sitter solution
 $a_i=( \sqrt{\frac{\rho_i+\mu^4}{3}\left(1-\frac{\rho_i+\mu^4}{\rho_c}\right)},\rho_i)$ satisfying
 $\omega=\omega_{eff}=-1$ and $\Omega=2$, to the de Sitter one  $a_f\cong(\frac{\mu^2}{\sqrt{3}},0)$ satisfying
 $\omega=\omega_{eff}=-1$ and $\Omega=0$.

\subsection{A teleparallel version of the $F(R)=\frac{R}{2}-\frac{\mu^4}{2R}$ model}
As a last example of teleparallel theory,  we consider  the function
 $G(T)=-\frac{T}{2}+\frac{3\mu^4}{2T}$, which corresponds to the model
$F(T)=\frac{T}{2}+\frac{\mu^4}{2T}$ (the teleparallel version of $F(R)=\frac{R}{2}-\frac{\mu^4}{2R}$).

In this case one has:
\begin{eqnarray} \label{20}
 \omega_{eff}=-1-\frac{f(\rho)}{\sqrt{\rho^2+3\mu^4}}, \quad \Omega=\frac{2\rho}{\rho+\sqrt{\rho^2+3\mu^4}}.
\end{eqnarray}

Then, for a universe filled by a pressureless fluid  $P=0$ (i.e., $f(\rho)=-\rho$), when $\rho$ takes large values one has
$\omega_{eff}\cong 0, \Omega\cong 1$, and for small values of $\rho$ one has $\omega_{eff}\cong -1, \Omega\cong 0$.
This means that, at early times ($\rho\rightarrow \infty$) the universe is in the matter dominated phase, and at the late times ($\rho=0$) it enters in
the de Sitter phase.
The universe  accelerates and is not singular at late times, but at early times it is singular. This could be easily
deduced because the universe is in the matter dominated phase at early times, or
from the conservation equation which for large values of $\rho$ reads $\dot{\rho}\cong -\sqrt{3}\rho^{3/2}$. Then, at early times, one gets
\begin{eqnarray}\label{21}
 H(t)\cong \frac{2}{3(t-t_s)},\quad \rho(t)\cong \frac{4}{3(t-t_s)^2}.
\end{eqnarray}

Finally we look for a  universe without singularities. Once again we consider the
EoS (\ref{15}) with $\rho_i\gg \sqrt{3}\mu^2$ which leads to a
 dynamical system with two critical points (two de Sitter solutions) at $a_f\equiv(\frac{\sqrt{3}\mu^2}{6},0)$ and
$a_i\equiv(\frac{\rho_i+\sqrt{\rho_i^2+3\mu^4}}{6},\rho_i)$. Note that, for this model, we do not need to introduce any small cosmological constant
because it is implicitly contained in the model.

At the de Sitter points $\omega_{eff}$ equals to $-1$ and $\Omega$ take the values
\begin{equation}\label{4.28}
 \Omega(a_f)=0 \mbox{ and } \Omega(a_i)=\frac{2\rho_i}{\rho_i+\sqrt{\rho_i^2+3\mu^4}}\cong 1,
\end{equation}
and when $\sqrt{3}\mu^2\ll \rho\ll\rho_i $, one has $\omega_{eff}\cong 0$ and $\Omega\cong 1$, that is,
the universe is in the matter dominated phase, which means that
the universe moves from the point $a_i$ to the point $a_f$ along the curve $\rho=-\frac{T}{2}+\frac{3\mu^4}{2T}$ passing trough a matter dominated phase.


The authors would like to thank Professor Sergei D. Odintsov for his valuable and useful comments.
This investigation has been
supported in part by MICINN (Spain), projects MTM2008-06349-C03-01 and MTM2009-14163-C02-02,
 and by AGAUR (Generalitat de Ca\-ta\-lu\-nya),
contracts 2009SGR 345 and 994.

\end{document}